\documentclass[aip,rsi,reprint,graphicx]{revtex4-1} 

\usepackage{lineno,hyperref}
\usepackage{graphicx,xcolor}
\usepackage{bm}

\modulolinenumbers[5]

\begin{document}

\title{A Sensitive Faraday Rotation Setup Using Triple Modulation}

\author{G. Phelps, J. Abney, M. Broering, W. Korsch}
\affiliation{Department of Physics and Astronomy, University of Kentucky, Lexington, KY 40506} 
\date{\today}

\begin{abstract}
The utilization of polarized targets in scattering experiments has become a common practice in many major accelerator laboratories.  Noble gases are especially suitable for such applications, since they can be easily hyper-polarized using spin exchange or metastable pumping techniques.  Polarized helium-3 is a very popular target because it often serves as an effective polarized neutron due to its simple nuclear structure.  A favorite cell material to generate and store polarized helium-3 is GE-180, a relatively dense aluminosilicate glass.  In this paper, we present a Faraday rotation method, using a new triple modulation technique, where the measurement of the Verdet constants of SF57 flint glass, pyrex glass, and air were tested.  The sensitivity obtained shows that this technique  may be implemented in future cell wall characterization and thickness measurements.  We also discuss the first ever extraction of the Verdet constant of GE-180 glass for four wavelength values of 632 nm, 773 nm, 1500 nm, and 1547 nm, whereupon the expected 1/$\lambda^{2}$ dependence was observed.    
\end{abstract}

\maketitle

\section{Introduction}

The study of spin interactions in scattering experiments has become a major part of modern accelerator laboratories.  The improvements on polarized noble gas targets, especially the generation of hyper-polarized helium-3, over the last few decades have been impressive. With the development of relatively narrow-band diode lasers, polarization values in excess of 50\% can be achieved at helium-3 densities of 8 amagats or more [\onlinecite{jaideep}].  In order to achieve such high polarization values, the choice of the cell material is a crucial component.  Typically a dense glass, such as GE-180, is used to store the gas in closed cell systems [\onlinecite{3He_cells}].  However, the maximally achievable polarization varies from cell to cell, even when identical fabrication and filling techniques are applied.  Thus, a better understanding and characterization of the glass thickness and magnetic properties are highly desirable. In this paper we propose how off-resonance Faraday rotation can be used to study such properties with high precision. 

Faraday showed that linearly polarized light would undergo a rotation of the plane of polarization upon being transmitted through a medium that has a magnetic field applied along the direction of propagation.  Therefore, a longitudinal magnetic field results in the medium becoming optically active [\onlinecite{schatz}].  In its simplest form the rotation, $\phi$, as expressed in Eq. (\ref{eq: faraday_effect}), is proportional to the strength of the magnetic field, $B_{0}$, and the length of the sample, \emph{l}: 

\vspace{-10pt}
\begin{equation} \label{eq: faraday_effect}
\phi =V\int^{l}_{0} B_{0} \,\mathrm dl^{\prime} =VlB_{0}.
\end{equation}

\noindent
\emph{V} is called the \emph{Verdet constant} which depends upon both the properties of the medium, the ambient temperature, and the wavelength, $\lambda$, of the incident light [\onlinecite{zvezdin}].  The sense of the angle of rotation, $\phi$, depends on the direction of the applied magnetic field, and by convention, \emph{V} is positive when the sense of the rotation is the same as the direction of the positive current that generates the magnetic field [\onlinecite{hougen}].  Achiral media are even under parity and odd under time-reversal, which distinguishes the Faraday magneto-optical effect from natural optical activity in chiral media [\onlinecite{buckingham}], [\onlinecite{chapter_4}].

Early quantum mechanical considerations of visible and ultraviolet light propagating through gaseous materials predicted a Verdet constant that varies approximately as the square of the frequency, where $\nu$ $\propto$ 1/$\lambda$.  Generally, the Faraday effect description reflected H. Becquerel's derived classical expression for the Verdet constant, $V$ = $(e\lambda/2mc)dn/d\lambda$, which shows that \emph{V} is proportional to the dispersion, $dn/d\lambda$ [\onlinecite{undergrad_lab}].  This describes a change in the index of refraction for left$-$handed versus right$-$handed circularly polarized light as a function of wavelength, where, in the long wavelength regime, the behavior of \emph{V} scales as $1/\lambda^{2}$. 

\section{System Setup}

Modulation techniques, in combination with lock-in amplifiers, serve as a powerful tool to extract small signals in a noisy environment.  A triple modulation method, with a noise floor of 60 nrad, has been devised, which allows for the extraction of small Faraday rotation angles in a single pass with very high precision.

\begin{figure}
\begin{center}
\includegraphics[height=35mm,width=85mm]{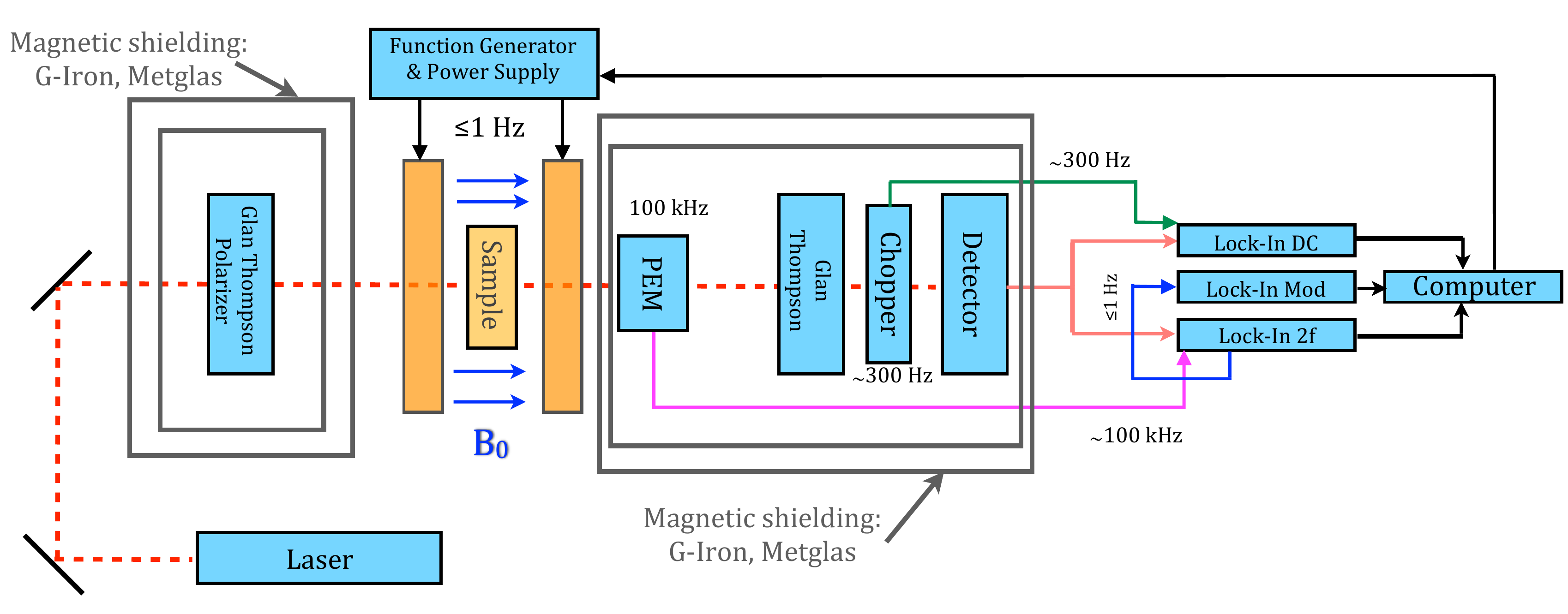}
\caption{Schematic of the triple modulation set-up, which includes a linear polarizer, Helmholtz coils with an applied sinusoidal magnetic field, PEM, second linear polarizer, optical chopper, and detector.  LabVIEW was used to control the magnetic field and to collect the data.  The G-Iron/Metglas magnetic shielding was implemented to eliminate any magnetic fringe fields.}
\label{fig: trip_mod}
\end{center}
\end{figure}

A schematic of the setup is shown in Figure (\ref{fig: trip_mod}).  Light is first incident upon a linear polarizer with the transmission axis oriented such that the emergent beam is \emph{s} polarized.  Then the light is transmitted through a sample that is in a magnetic field generated by a pair of Helmholtz coils.  Upon transmission, the light then goes through the polarization state analyzer, which is comprised of a photoelastic modulator (PEM) and second linear polarizer with transmission axis at 45$^\circ$ w.r.t both the optical axis of the PEM and transmission axis of the first polarizer.  Finally, an optical chopper is placed in front of the detector.  Due to the magnetic fringe fields from the Helmholtz coils, components away from the Helmholtz coils were wrapped with two layers of shielding.

Three parameters are modulated in our set-up: (i) the magnetic field, (ii) the light polarization, and (iii) the light intensity.  A Wavetek 81 function generator is used to drive the current supply to the Helmholtz coils, which generates a sinusoidal magnetic field at a low frequency of 1 Hz or below.  To dynamically polarize the light, a Hinds photoelastic modulator, with a nominal frequency of 50 kHz, is implemented.  For the minimization of drifts in the DC signal, a Thorlabs MC2000 optical chopper, with an operating frequency of 300 Hz, is placed after the polarization state analyzer.  Utilizing the Stokes vector formalism, the intensity at the detector is given by the following expression:

\vspace{-10pt}
\begin{equation} \label{eq: intensity}
I=I_{0}\left(1-4J_{1}(A)\epsilon(t)\sin(\omega t)+4J_{2}(A)\phi(t)\cos(2\omega t)\right),
\end{equation}

\noindent
where \emph{A} is the Bessel angle of the PEM, $J_{i}(A)$ results from the Bessel angle expansion of the retardation of the PEM, $\epsilon$ represents the ellipticity, and $\phi$ is the rotation.  The quantity \emph{A} is typically chosen to be 2.405 radians, such that $J_{1}(A) = 0.519$, and $J_{2}(A)=0.432$.  Eq. (\ref{eq: intensity}) may be divided into three terms: the overall scaling factor, $I_{0}$, is proportional to the laser intensity; a term that is at the frequency of the PEM (50 kHz), denoted as the 1$\omega$ term, is related to the ellipticity; and a quantity at twice the frequency of the PEM (100 kHz), referred to as the 2$\omega$ term, gives rise to the rotation.  

\section{Signal Analysis}
The light detector signal is modulated three-fold, therefore, three lock-in amplifiers (Signal Recovery, model 7265) were used to extract the desired signal.  One lock-in amplifier is referenced to the frequency of the optical chopper with an integration time of 500 ms; this signal is given by $I_{0}$ in Eq. (\ref{eq: intensity}).   A second lock-in amplifier uses the 100 kHz signal from the PEM for detection of the rotation, where an integration time of 320 $\mu$s was used; however, the sinusoidally driven magnetic field results in an additional time dependence of the rotation.  Since the magnetic field is modulated at a slow rate of 1 Hz or less, this component of the signal essentially is not effected by the 100 kHz lock-in amplifier, such that it can be passed through the device for further processing.  A third lock-in amplifier was used to detect this slowly varying signal, where integration time constants up to 1000 s were applied, depending upon the magnitude of the magnetic field amplitude and the Verdet constant of the sample.

Care must be taken when extracting the Faraday rotation signal, so that all of the correct conversion factors are included.  The rear output of the 100 kHz lock-in amplifier generates a voltage signal that is scaled as a percentage of 10 V, based upon the percentage of the front panel voltage value to the sensitivity setting.  Therefore, one may express the real signal in terms of the 100 kHz lock-in amplifier's front panel display, $V^{LI}$, as:

\vspace{-10pt}
\begin{equation}
V^{real}_{mod}=\frac{2\cdot\sqrt[]{2}\cdot\sqrt[]{2}}{10 V/G^{LI}}V^{LI}_{mod}=\frac{2\cdot G^{LI}}{5 V}V^{LI}_{mod}.
\end{equation}

\noindent
In the above expression, $G^{LI}$ represents the sensitivity setting on the front panel of the 100 kHz lock-in amplifier in units of Volts.  The factor of 2 is a result of the optical chopper, which decreases the overall signal by 2.  Also, the front panel voltage of lock-in amplifiers display the rms value of the signal instead of the peak value.  Therefore, to recover the modulated rotation signal, two factors of \hspace{1pt} $\sqrt[]{2}$ \hspace{1pt} are included, since two lock-in amplifiers are used.  This was verified by measuring the output voltages on an oscilloscope.

Furthermore, the signal of the 300 Hz lock-in amplifier is affected by the square wave signal of the optical chopper, such that a Fourier decomposition of the signal is needed.  The displayed voltage on this lock-in amplifier multiplies the real signal by a factor of 2/$\pi$, with the recovery of the peak voltage signal, here denoted as $I_{0}$,  being expressed as:

\vspace{-10pt}
\begin{equation}
I_{0}=\frac{\sqrt[]{2}}{2} \pi I^{LI}_{0}.
\end{equation}

\noindent
Once again the additional factor of \hspace{1pt} $\sqrt[]{2}$ \hspace{1pt} comes from the lock-in amplifier displaying the rms value.  Therefore, upon normalizing to $I_{0}$, the ratio of 2$\omega$/$I_{0}$ to isolate the rotation signal, becomes:

\vspace{-10pt}
\begin{equation}
4J_{2}(A)\phi(t)=\frac{4\hspace{2pt}\sqrt[]{2}\cdot G^{LI}}{10\pi}\frac{V^{LI}_{mod}}{I^{LI}_{0}},
\end{equation}

\noindent
such that the Faraday rotation is obtained to be:

\vspace{-10pt}
\begin{equation}
\phi(t)=\frac{\sqrt[]{2}\cdot sens}{10\pi J_{2}(A)}\frac{V^{LI}_{mod}}{I^{LI}_{0}}=VlB_{0},
\end{equation}

\noindent
where \emph{V} is the Verdet constant, \emph{l} is the effective length of the sample, and $B_{0}$ is the amplitude of the applied magnetic field, which was measured using a Lakeshore 475 Gaussmeter with an axial probe [\onlinecite{gretty_thesis}].

\section{Experimental Results}
Three Thorlabs lasers were used in mapping \emph{V} vs. $\lambda$ for four wavelength values: a $\lambda$ = 632 nm frequency stabilized HeNe, and two external cavity Littman configuration tunable diode lasers with nominal wavelengths $\lambda$ = 770 nm and $\lambda$ = 1550 nm.  The wavelengths probed were 632 nm, 773 nm, 1500 nm, and 1547 nm, where a Burleigh WA-1500 wavemeter was used to measure the wavelength.  For measurements made at $\lambda$ = 632 nm and $\lambda$ = 773 nm, a Hinds DET-100 photodiode detector was used, where the photo detector had an active area of 16 mm$^{2}$.  A Hinds DET-200 photodiode detector, with an active area of 8 mm$^{2}$, was used for measurements at 1500 nm and 1547 nm.  Furthermore, all measurements were performed at standard pressure and temperature.

\subsection{Calibration Glass Measurements}
In order to test the sensitivity of the apparatus, measurements on samples with well-characterized Verdet constants were performed. As test cases, flint glass SF57 and pyrex glass were used.
SF57 flint glass has a known Verdet constant of $2.01\times 10^{-5}$ rad/(G$\cdot$cm) at 300 K and $\lambda = 632.8$ nm [\onlinecite{weber}].  Samples of approximately 1/2$^{\prime\prime}$ $\times$ 1/2$^{\prime\prime}$ $\times$ 1/2$^{\prime\prime}$ were obtained from Glass Fab Inc., and were optically polished at Dell Optics Company.  The first measurement for \emph{V} was at $\lambda$ = 632 nm, resulting in \emph{V} = (2.0259 $\pm$ 0.0004 $\pm$ 0.005) $\times \hspace{3pt}10^{-5}$ rad/(G$\cdot$cm), which agrees nicely with [\onlinecite{weber}].  Upon investigating at longer wavelengths, the data shows that the expected 1/$\lambda^{2}$ dependence is apparent.  The data appears in Table (\ref{tab: v_sf57}), where the quoted errors are the statistical and systematic, respectively.

\begin{table}
\centering
\begin{tabular}{|c|c|c|}
\hline
$\lambda$ \hspace{1pt}(nm) & V$_{SF57}$ \hspace{1pt}($\times 10^{-5}$rad/(G$\cdot$cm)) & V$_{pyrex}$ \hspace{1pt}($\times 10^{-6}$rad/(G$\cdot$cm))\\
\hline
632 & 2.0259 $\pm$ 0.0004 $\pm$ 0.005 & 3.507 $\pm$ 0.004 $\pm$ 0.08 \\
\hline
773 & 1.13 $\pm$ 0.01 $\pm$ 0.003 & 2.3 $\pm$ 0.1 $\pm$ 0.05 \\
\hline
1500 & 0.239 $\pm$ 0.005 $\pm$ 0.0007 & 0.51 $\pm$ 0.05 $\pm$ 0.01 \\
\hline
1547 & 0.2298 $\pm$ 0.0006 $\pm$ 0.0006  & 0.48 $\pm$ 0.03 $\pm$ 0.01 \\
\hline
\end{tabular}
\caption{The measured Verdet constant of SF57 flint glass and pyrex glass for various $\lambda$, with air corrections applied.}
\label{tab: v_sf57}
\end{table}

Pyrex glass has also been studied in previous experiments, with $V_{pyrex} \approx 3.64 \times 10^{-6}$ rad/(G$\cdot$cm) for $\lambda = 632$ nm [\onlinecite{pyrex}].  A 4.7 mm thick sample was implemented as a second calibration sample in testing the triple modulation method.  For $\lambda$ = 632 nm, we found $V_{pyrex} = (3.507 \pm 0.004 \pm 0.08)\times 10^{-6}$ rad/(G$\cdot$cm), in agreement with with [\onlinecite{pyrex}].

\subsection{Contribution Due to Air}

There have been various experiments probing the Faraday effect of air, where reports for $V_{air}$ have ranged from ($1.3 \times 10^{-9} $\hspace{2pt}$-$ $1.9 \times 10^{-9})$ rad/(G$\cdot$cm), with the most recent experiment from 2011 measuring $1.4 \times 10^{-9}$ rad/(G$\cdot$cm) [\onlinecite{jacob}], [\onlinecite{lin}], [\onlinecite{cy}].   Air has a small contribution to the Faraday rotation of the measured glass samples that must be subtracted.  To measure the contribution due to air using the triple modulation method, a magnetic field with amplitude B$_{0}\le$ 61 G was used.  In order to improve the signal-to-noise ratio, an integration time constant of 1000 s was implemented on the lock-in amplifier detecting the desired signal from the field modulation.  A value of V$_{air} = (1.414 \pm 0.003 \pm 0.004) \times 10^{-9}$ rad/(G$\cdot$cm) was extracted, which agrees well with the references [\onlinecite{lin}], [\onlinecite{cy}]. The measured wavelength dependence of V$_{air}$ is displayed in Table(\ref{tab: v_air}), and these values were subtracted off as small corrective terms when measuring the Verdet constant of the glass samples.  

Furthermore, tests were performed to extract the noise floor of the triple modulation method via measuring the Faraday rotation of air for various applied magnetic field amplitudes.  These amplitudes extended to 0.05 G, whereupon a value of 60 nrad was established for this technique.  As the precision of these results indicate, the triple modulation method for measuring Faraday rotations provides a sensitive apparatus that may be employed in determining Verdet constants of unknown materials, as well as distinguishing variant chemical compositions that may arise in the fabrication process of these materials.

\begin{table}
\centering
\begin{tabular}{|c|c|}
\hline
$\lambda$ \hspace{1pt}(nm) & V$_{air}$ \hspace{1pt}($\times10^{-9}$ rad/(G$\cdot$cm)) \\
\hline
632 & 1.414 $\pm$ 0.003 $\pm$ 0.004 \\
\hline
773 & 0.8 $\pm$ 0.2 $\pm$ 0.003 \\
\hline
1500 & 0.27 $\pm$ 0.04 $\pm$ 0.0008 \\
\hline
1547 & 0.12 $\pm$ 0.02 $\pm$ 0.0003  \\
\hline
\end{tabular}
\caption{Measured Verdet constant of air for various $\lambda$.}
\label{tab: v_air}
\end{table}

\section{GE-180 Glass}

To aid in making the first ever measurement of the Verdet constant of GE-180, a approximately 4 mm thick circular plate, borrowed from NIST, was employed.  The Verdet constant for GE-180 is unknown; therefore, the thickness of the plate needs to be exactly known so that a precise measurement of the Verdet constant may be obtained.  This was accomplished by utilizing an interferometry technique with a New Focus tunable diode laser, where the thickness determination was (4.00 $\pm$ 0.02) mm [\onlinecite{gretty_thesis}].

\begin{figure}
\begin{center}
\includegraphics[height=55mm,width=85mm]{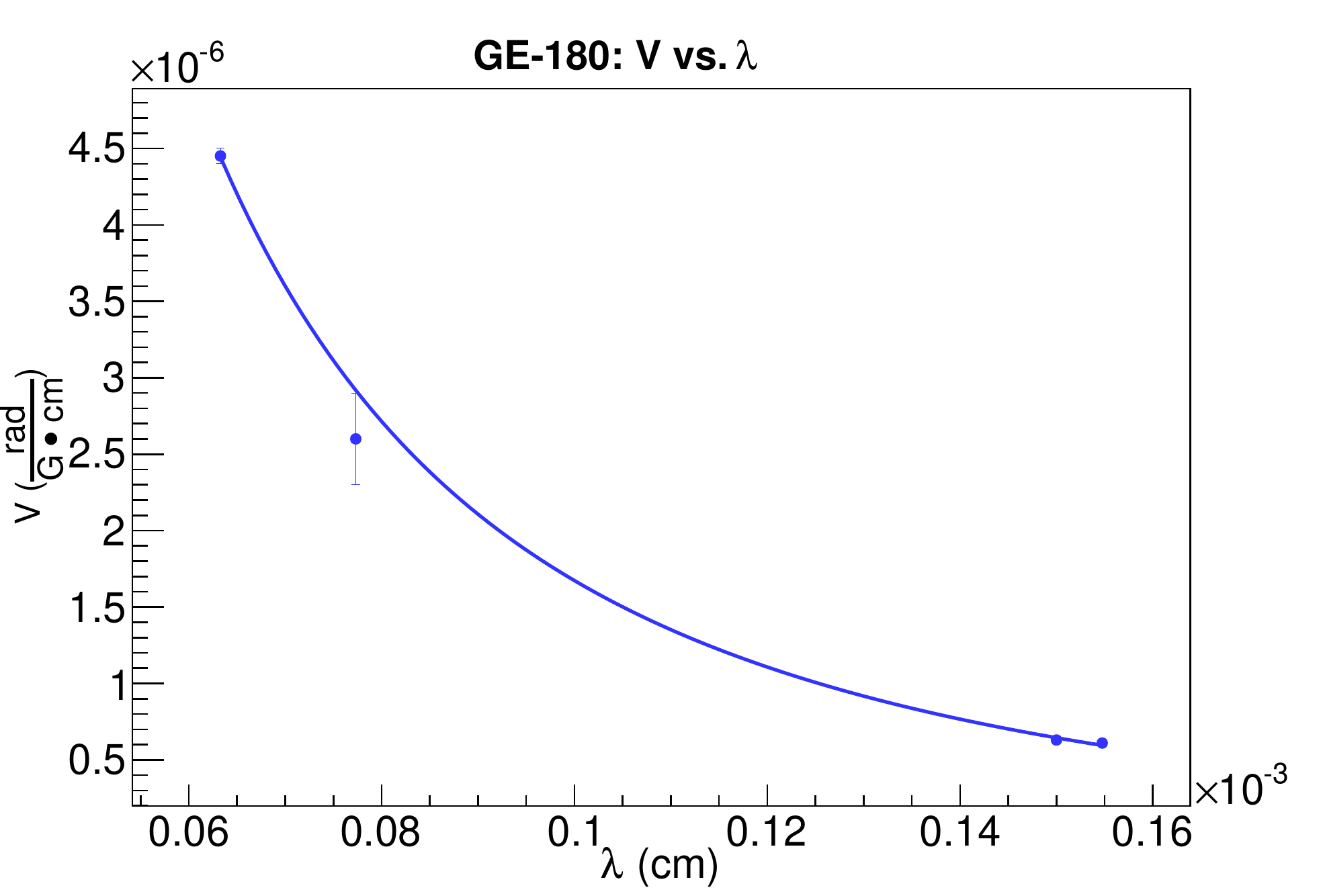}
\caption{Plot of the Verdet vs. $\lambda$ for the 4 mm GE-180 sample from NIST.}
\label{fig: GE-180_data}
\end{center}
\end{figure}

Additionally, with a precise measurement for the thickness of the NIST GE-180 sample, a study of the Verdet constant with respect to wavelength was conducted.  Table (\ref{tab: v_GE-180}) displays the results obtained for the first study of the Faraday effect for GE-180.  Furthermore, the expected 1/$\lambda^{2}$ dependence is shown in Fig. \hspace{-3pt}(\ref{fig: GE-180_data}).

\begin{table}
\centering
\begin{tabular}{|c|c|}
\hline
$\lambda$ \hspace{1pt}(nm) & V$_{GE-180}$ \hspace{1pt}($\times10^{-6}$ rad/(G$\cdot$cm)) \\
\hline
632 & 4.451 $\pm$ 0.004 $\pm$ 0.01 \\
\hline
773 & 2.6 $\pm$ 0.3 $\pm$ 0.007 \\
\hline
1500 & 0.63 $\pm$ 0.02 $\pm$ 0.002 \\
\hline
1547 & 0.61 $\pm$ 0.02 $\pm$ 0.005  \\
\hline
\end{tabular}
\caption{The measured Verdet constant of the NIST GE-180 sample for various $\lambda$, with the air corrections applied.}
\label{tab: v_GE-180}
\end{table}

The knowledge of $V_{GE-180}$ opens up new possibilities for measuring the wall thicknesses of glass cells with curved surfaces.  Often the GE-180 target cells used for polarized helium-3 scattering experiments are composed of spherical or cylindrical components, where optical interference methods for the determination of the wall thickness are difficult to apply.  In such experiments, the amount of material which is traversed by the particles needs to be known 
in order to properly correct for radiative and multiple scattering effects.  However, once the Verdet constant is extracted from a calibration sample, the modulation technique described in this letter can be applied to the measurement of the wall thicknesses.  If the cell is exposed to a uniform magnetic field, the Faraday rotation probes the total thickness of the two walls that the light traverses. However,  subjecting the front and back wall to different fields, e.g. by generating a field gradient across the sample, the contributions from the two walls can be determined separately.  

An additional GE-180 sample was obtained from Princeton University and the Verdet constant was extracted at $\lambda$ = 632 nm. This measurement yielded a result of $V_{Princeton}$ = (3.790 $\pm$ 0.007 $\pm$ 0.050) $\times$ 10$^{-6}$ rad/(G$\cdot$cm), which is approximately 20$\%$ lower than the NIST GE-180 sample.  The difference between the two extracted values is larger than the determined uncertainties.  This may be attributed to a difference in manufacturing procedures or in the chemical composition between the two GE-180 samples. It should be noted that the origin of the two samples could not be reconstructed.

\section{Conclusion}

A sensitive Faraday rotation setup with a noise floor of 60 nrad has been developed. The system is based on a triple modulation technique using a single-path laser beam. The Verdet constant of the aluminosilicate glass GE-180 was extracted for the first time.  It was found that two different calibration samples yielded two slightly different values for V$_{GE-180}$, an indication of a minor difference in chemical composition or manufacturing process.  GE-180 is a very popular material for the production of dense polarized helium-3 targets, which are often utilized in modern low, medium, and high energy scattering experiments. The technique described here may usher in a new procedure for diagnosing such target cells where the precise knowledge of the cell wall thicknesses, relevant information for radiative corrections and the understanding of multiple scattering effects, is important.

\begin{acknowledgements} 
This work was supported by the Department of Energy award no. DE-FG02-99ER41101.  The authors would also like to thank T. Gentile (NIST) and M. Souza (Princeton University) for providing the GE-180 test samples.
\end{acknowledgements}

\end{document}